# On forced RF generation of CW magnetrons for SRF accelerators

G. Kazakevich[#], R.P. Johnson, Muons, Inc, Newport News, VA 23606, USA;
T. Khabiboulline, G. Romanov, V. Yakovlev, Fermilab, Batavia, IL 60510, USA;
Ya. Derbenev, JLab, Newport News, VA 23606, USA;
Yu. Eidelman, Eidelman Scientific Consulting, Naperville IL 60563, USA.

*Abstract*

CW magnetrons, initially developed for industrial RF heaters, were suggested to power RF cavities of superconducting accelerators due to their higher efficiency and lower cost than traditionally used klystrons, IOT's or solid-state amplifiers. RF amplifiers driven by a master oscillator serve as coherent RF sources. CW magnetrons are regenerative RF generators with a huge regenerative gain. This causes regenerative instability with a large noise when a magnetron operates with the anode voltage above the threshold of self-excitation. Traditionally for stabilization of magnetrons is used injection locking by a quite small signal. Then the magnetron except the injection locked oscillations may generate noise. This may preclude use of standard CW magnetrons in some SRF accelerators. Recently we developed briefly described below a mode for forced RF generation of CW magnetrons when the magnetron startup is provided by the injected forcing signal and the regenerative noise is suppressed. The mode is most suitable for powering high Q-factor SRF cavities.

## 1. Introduction

High-power CW magnetrons, designed and optimized for industrial RF heaters, but driven by an injection-locking signal, were suggested in number of works to power Superconducting RF (SRF) cavities in accelerators due to higher efficiency and significantly lower cost of generated RF power per Watt than traditionally used RF amplifiers (klystrons, IOT's, solid-state amplifiers). The RF amplifiers driven by a master oscillator serve as coherent low noise RF sources. The CW magnetrons are regenerative RF generators with a huge regenerative gain of the resonant system to start up reliably with a self-excitation by noise even if the tube is powered by a DC power supply. Very large regenerative gain causes a regenerative instability with a large regenerative noise. Traditionally the magnetrons operate in the self-excitation mode, i.e. with the anode voltage above the self-excitation threshold, with a small injection-locking signal, $P_{Lock} \approx$ -20 dB or less of the magnetron power $P_{Mag}$. In this case the regenerative noise of a CW magnetron violates a necessary correlation of the tube startup with its injection-locking, i.e., the magnetron may be launched by the noise, but not by the injection-locking signal. Such probability is considered in the presented work. A developed method of forced RF generation CW magnetrons eliminating startup by noise and PS ripples is briefly described below. The method was verified in experiments with CW magnetrons of microwave ovens.

## 2. Operation of a CW injection-locked magnetron in the self-excitation mode

We consider operation of a CW magnetron as it is traditionally assumed, in the self-excitation mode, at a injection-locking signal with low power $P_{Lock}$.

The effective bandwidth of injection-locking $\Delta f$, at the locking signal is expressed by the following equation [1]:

$$\Delta f = \frac{f_0}{2Q_L} \sqrt{\frac{P_{Lock}}{P_{Mag}}} \ . \qquad (1)$$

Here $f_0$ is the instantaneous magnetron frequency, $Q_L$ is the magnetron loaded Q-factor, $P_{Mag}$ is the magnetron output power. For the 2.45 GHz, free running microwave oven magnetron type 2M137-IL the effective bandwidth $\Delta f_{FR} \approx$ 4.5 MHz [2]. Out of the effective bandwidth the magnetron cannot be injection-locked at the given $P_{Lock}$.

Then the probabilities of the injection-locking process $w_{Lock}$ and a free running operation $w_{FR}$ ($P_{Lock}$ =0) for 2.45 GHz, CW tube one estimates as:

$$w_{Lock} \sim \frac{\Delta f}{\Delta f_{FR}} \ . \qquad (2)$$

$$w_{FR} \sim \frac{\Delta f_{FR} - \Delta f}{\Delta f_{FR}} \ . \qquad (3)$$

The probabilities estimates vs. $P_{Lock}$ are shown in Table 1.

Table 1. The values of $w_{Lock}$ and $w_{FR}$ vs. $P_{Lock}$.

| $P_{Lock}$ | $\Delta f$, | $w_{Lock}$ | $w_{FR}$ |
|---|---|---|---|
| -10 dB | 3.87 MHz | ~0.86 | ~0.14 |
| -20 dB | 1.22 MHz | ~0.27 | ~0.73 |
| -30 dB | 0.39 MHz | ~0.09 | ~0.91 |

Thus, probability of the injection-locked generation of such RF source may be quite low, notably less than probability of the free running generation caused by noise. The noise oscillations are much less in magnitude than the injection-locked those. This leads to a notable quasicontinuous noise spectrum which is a disadvantage for operation of the self-exciting magnetrons with low locking signal for high Q-factor SRF cavities. The distorted spectra of the RF sources with an intense quasi-continuous noise may preclude required suppression of parasitic modulations (microphonics, etc.) and may increase emittance of the beam in SRF accelerators. Note, if the effective bandwidth of the injection-locking is approached to zero, i.e., the transient time of the injection locking tends to infinity [1], then probability of the injection-locking also approaches to zero.

[#] gkazakevitch@yahoo.com, grigory@muonsinc.com

We have studied applicability of CW magnetrons for various projects of SRF accelerators. Experimental verifications were performed with 2.45 GHz microwave oven magnetrons. Figure 1, [3], shows measured spectral density of the noise power relatively the carrier frequency $f_c$ vs. the power of the injection-locking signal $P_{Lock}$ for the magnetron type 2M137-IL operating above the self-excitation threshold voltage.

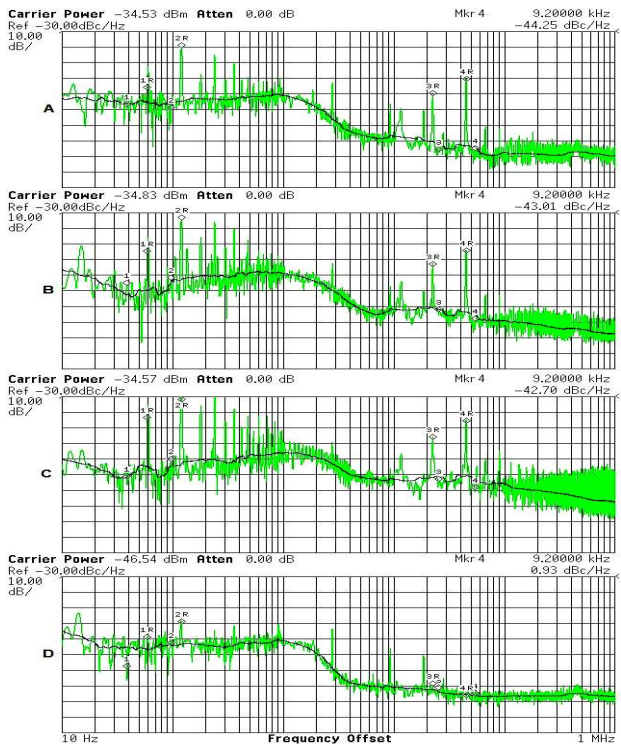

Fig. 1. The spectral power density of noise relatively $f_c$ of the magnetron 2M137-IL at the output power of 1 kW, at the locking signal of 100, 30, and 10 W, traces A, B, and C, respectively. Traces D are the spectral power density of noise of the injection-locking signal ($P_{Lock}$ =100 W), when the magnetron feeding voltage is OFF. Black traces are the averaged spectral power density of the noise.

Measured in Fig. 1 a significant (by ~20 dBc/Hz) increase of the spectral density of noise power ($|f_{Lock} - f_0|$ >100 kHz) one can explain as quite narrow bandwidth of the injection-locking process and its low probability for the magnetron at the locking signal of -20 dB (10 W).

## 3. Operation of CW magnetron above and below the self-excitation threshold voltage

Measured with bandwidth resolution of 5 Hz, Fig. 2, the carrier frequency offset shows quasicontinuous noise spectra of the magnetron type 2M137-IL when the tube operated above or below the self-excitation threshold voltage at various injected resonant signal power [3].

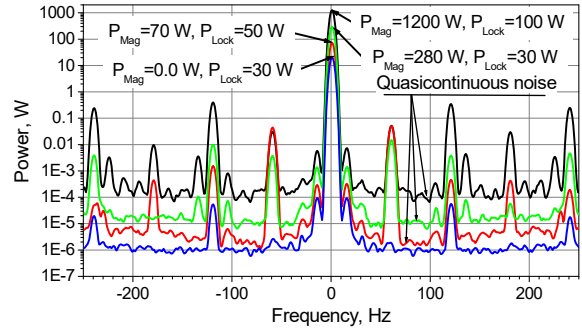

Fig. 2. Offset of the carrier frequency of the magnetron type 2M137-IL (operating in CW mode) with the threshold of self-excitation of 4.04 kV at various power levels of magnetron output, $P_{Mag}$, and the injection-locking (forcing) signal, $P_{Lock}$.

The trace $P_{Mag}$ =0.0 W, $P_{Lock}$ =30 W shows the carrier frequency offset of the injection-locking signal when the magnetron anode voltage was OFF. The traces $P_{Mag}$ =70 W, $P_{Lock}$ =50 W and $P_{Mag}$ =280 W, $P_{Lock}$ =30 W show the tube operation below the self-excitation threshold voltage. The trace $P_{Mag}$ =1200 W, $P_{Lock}$ =100 W relates to operation above the self-excitation voltage. The tube anode voltages $U_{Mag}$ in this experiment were 3.90, 4.01 and 4.09 kV, respectively.

Traces in Fig. 2 show that the ratio of carrier frequency peak to quasicontinuous noise is largest at operation below the self-excitation voltage. For high Q cavities the most suitable operation is below the self-excitation threshold voltage with a large resonant injected signal.

Presently the theory of magnetrons operation is absent; the existing phenomenological models of magnetrons operation proposed 60-80 years ago cannot provide capabilities to choose the CW magnetrons design and parameters for most suitable application of the tubes in various SRF accelerators projects.

Using the model of the charge drift approximation [4] and equations for motion of electrons in a cylindrical magnetron [5], we found the necessary and sufficient conditions for startup and operation of magnetrons, taking into account the phase grouping of Larmor electrons in the "spokes" [3] for the tubes fed even below the Hartree voltage. Considering RF generation of magnetrons as non-stationary processes, makes it possible to find the mode of operation and control of magnetrons as coherent RF sources with maximum efficiency, phase and power controllability in a wide frequency band, with a much lower the spectral density noise power [6].

Basing on experiments with microwave oven magnetrons we demonstrate here applicability of the CW magnetrons operating in the Stimulated RF generation mode [7] for various SRF accelerator projects.

RF generation of a CW magnetron one can consider as a wide-band spontaneous emission in the interaction space amplified by a resonant system [6]. The magnetron total noise power depends on the regenerative noise of the resonant system $P_{RN}$ and the noise of the spontaneous emission $P_{SN}$, causing the quasicontinuous noise. The last one depends on phase grouping in a magnetron that, in turn, depends on the azimuthal component of electric field of the synchronous wave rotating in the space of interaction of the tube [6]. An increase of the injected resonant signal driving a magnetron increases the azimuthal electric field of the synchronous wave that improves the phase grouping of charges in spokes.

Figure 3 shows total noise power $P_{RN}+P_{RS}$, excluding sidebands vs. the tube anode voltage at various injected resonant signals and the output carrier peak power $P_{Mag}$ as it is plotted in Fig. 2 [6]. It is clearly seen that reducing the magnetron anode voltage allows significantly reduce the tube total noise.

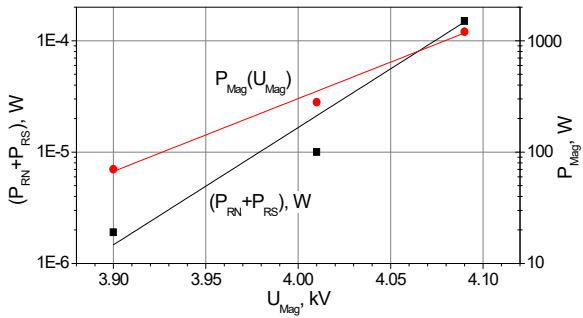

Fig. 3. Dependences of the magnetron total noise power (black dots, left scale) and the output power (red dots, right scale) vs. the anode voltage. Solid lines show the exponential fits of build-ups of the output power, $P_{Mag}$ and $P_{RN}+P_{RS}$ power for the magnetron 2M137-IL [6].

Affect of the phase grouping on the $P_{RS}$ was studied measuring carrier frequency offset nearby the threshold of self-excitation [8] at the anode voltage by ≈150 V less than the threshold of self-excitation at RBW=100 kHz.

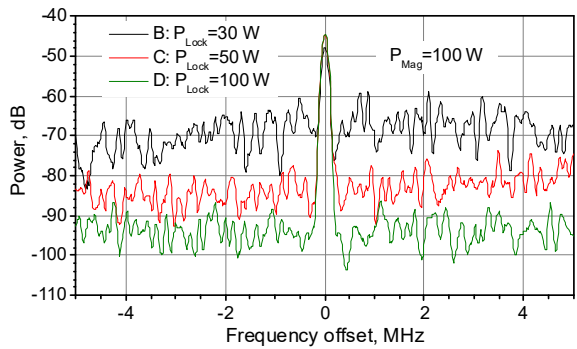

Fig. 4. Offset of the carrier frequency of magnetron type 2M137-IL measured at the magnetron output power 100 W vs. power of the injection-locking signal, $P_{Lock}$ [8].

Smoothed by method of adjacent averaging, the offset traces vs. the injected resonant signal are shown in Fig 5.

It is seen that the injected signal at $P_{Lock}$ =30 W cannot launch the magnetron with this anode voltage. The trace shows the spontaneous emission caused by motion of charges in the interaction space and amplified by the magnetron resonant system.

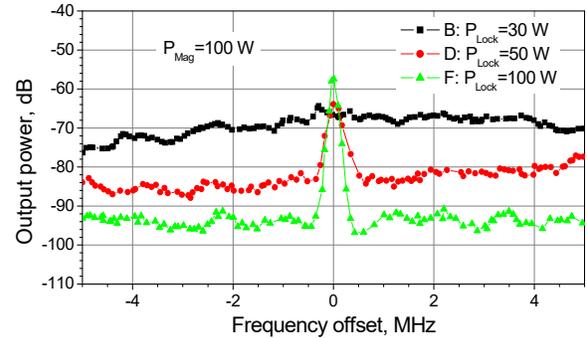

Fig. 5. Smoothed offset of the carrier frequency of the 2M137-IL magnetron at the magnetron output power 100 W vs. power of the injection-locking signal, $P_{Lock}$ [6].

In this figure the trace at $P_{Lock}$ =30 W indicates the wide-band oscillation in the interaction space at the bandwidth of ≈7 MHz at the level of -3 dB amplified by a resonant system up to ~1 W output power. One can imply that this trace shows the amplified spontaneous radiation close to point of launching i.e., in fact, the quasicontinuous noise. Other traces show the incoherent spontaneous radiation significantly reduced due to its conversion by the phase grouping into the coherent generation when the magnetron is launched [6]. Thus, the spontaneous radiation becomes mostly coherent one (phase-locked) due to the phase grouping. Its residual power $P_{RS}$ characterizes the loss of coherency at the phase grouping vs. $P_{Lock}$.

A noticeable frequency shift of the phase-locked spontaneous oscillations at $P_{Lock}$ =50 W indicates the frequency pushing of the oscillations caused by the anode current of launched magnetron. A further increase of the locking signal reduces the intensity $P_{RS}$ of spontaneous oscillations, converting them mainly into coherent ones.

## 4. Stimulated generation mode of a CW magnetron

We have developed the Stimulated generation mode and tested it experimentally in pulse regime using a CW 2.45 GHz magnetron type 2M219G with nominal output power of 945 W and the measured magnetron self-excitation threshold voltage of 3.69 kV [7]. The mode uses operation of a magnetron below the threshold of self-excitation, with quite large injected forcing signal, but the anode voltage is chosen to eliminate any startups of the tube by noise or power supplies ripples. Main characteristics of this mode are briefly presented below.

Solid lines in Fig. 6 show the ranges of power control in the CW magnetron vs. the magnetron anode voltage and the injected resonant signal power. The magnetron was powered by a pulse High Voltage (HV) source using a partial discharge of the 200 μF storage capacitor, [9], providing pulse duration of ≈5 ms. The pulsed HV source was powered by a charging Glassman 10 kV, 100 mA switching power supply allowing for voltage control.

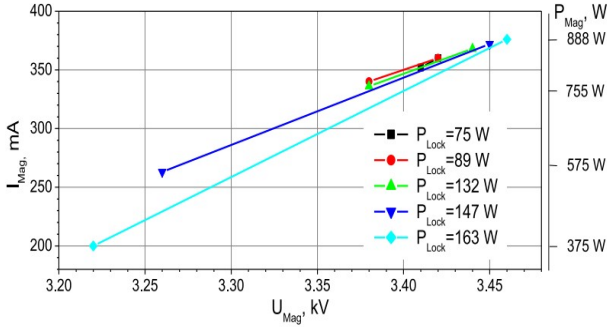

Fig. 6. The ranges of the 2M219G magnetron anode voltage and the magnetron current in the Stimulated generation mode at various power levels of the injected resonant signal $P_{Lock}$. The right scale shows measured RF output power of the magnetron corresponding to the magnetron current, $I_{Mag}$ [7].

As it follows from Fig. 6, the injected resonant signal of -7.6 dB (163 W) allows the range of the magnetron power control of ≈ 3.7 dB by variation of the magnetron current in this tube. An increase of the range of the tube power (current) control is caused by larger coherent gain [6] for charges moving in spokes due to improved phase grouping in the space of interaction at larger injected resonant forcing signal.

The magnetron performs pulse forced Stimulated RF generation by injection of a pulse forcing resonant signal into the magnetron, Figs. 7, 8 at a DC feeding power.

Despite the significantly lower regenerative gain at a reduced anode voltage, the magnetron in the Stimulated generation mode with a large forcing signal converts the spontaneous oscillations in the interaction space mainly into coherent ones [6], which are amplified by the resonant system of the tube. This provides almost rated output coherent RF power at the large coherent gain [6].

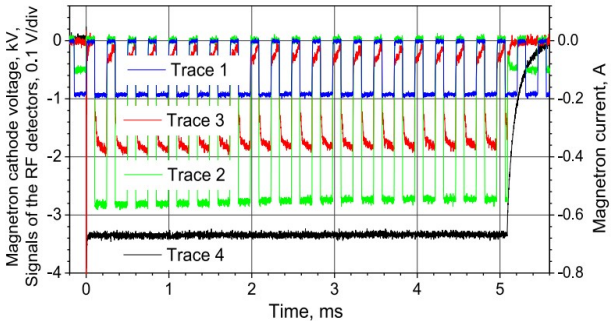

Fig. 7. 4 kHz trains of 147 μs pulses (duty factor of ≈59%). Traces 1 and 2 - are the resonant injected and the magnetron output RF signals with powers of 125 W and 803 W, respectively; trace 3 - is the magnetron pulse current (right scale); trace 4 - is the magnetron cathode voltage (−3.37 kV). The magnetron pulse current was measured by a current transducer (type LA 55-P) with a circuit integration time ≈50 μs.

The absence of noise at the absence of a forcing resonant injected signal in the Stimulated generation mode is represented by the traces in Figs. 7 and 8 [7].

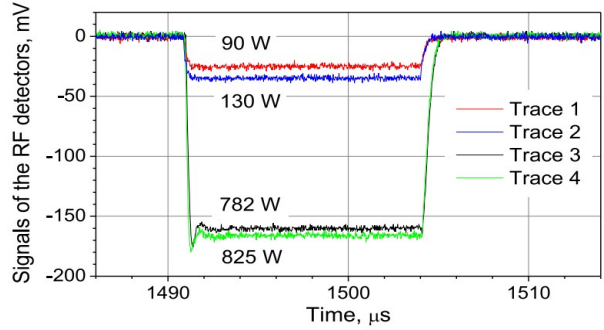

Fig. 8. Measured pulses of the 13 μs 20 kHz train when the magnetron operates in the Stimulated generation mode. Traces 3 and 4 are the magnetron output RF signals in dependence on the power of driving signals shown in traces 1 and 2, respectively.

The conversion efficiency $\eta$, expressed by the ratio of the generated RF power $P_{RF}$ to the consumed power of a CW magnetron operating in the Stimulated generation mode, neglecting the filament power, is determined by the following expression:

$$\eta \approx P_{RF}/(U_{Mag} \cdot I_{Mag} + P_{Lock}) \qquad (4)$$

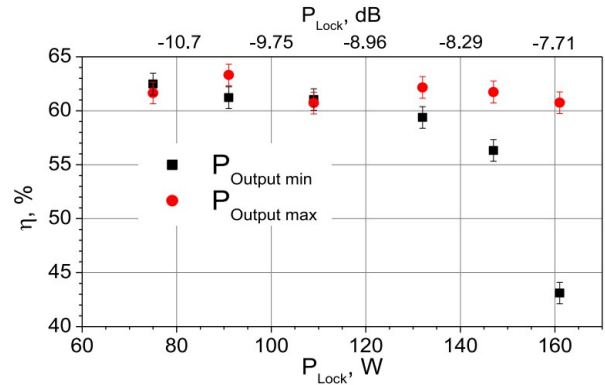

Fig. 9. Dependence of conversion efficiency of the 2M219G magnetron on power of injected signal $P_{Lock}$ [7].

The plotted measured results relate to the maximum and minimum power of the magnetron operating in the Stimulated generation mode, showing the range of the magnetron power control vs. $P_{Lock}$, varying the magnetron current. This method of power control in a wide range is most efficient [8]. The measured conversion efficiency of the magnetron operating above the self-excitation threshold in the "free run" mode ($U_{Mag} \approx 3.69$ kV, $P_{Lock} = 0$) at the nominal tube power is ≈ 54%.

Usage of two-cascade magnetrons allows reducing the forcing signal for Stimulated generation mode to ≈ -20 dBc [3]. In this scheme both magnetrons operate in the Stimulation generation mode; the first one with power ≈10% of the required RF power drives the second one realizing the required power control by variation of the magnetron current. The cost of two-cascade magnetrons in mass production should be much lower than the cost of traditionally used RF sources.

## 5. Bandwidth of phase and power control of magnetrons in Stimulated generation mode

For various SRF accelerator projects (colliders, ADS facilities and even industrial SRF accelerators is important wide-band control of SRF sources in phase and power for suppression various parasitic modulations (microphonics, etc.). The Stimulated generation mode utilizes quite large injected resonant signal necessary for such control; it is most suitable for magnetron RF sources in various SRF accelerators.

The bandwidth of control $BW_C$ for 2.45 GHz microwave oven magnetrons obtained measuring the transfer functions magnitude and phase characteristics [3] is presented in Fig. 10.

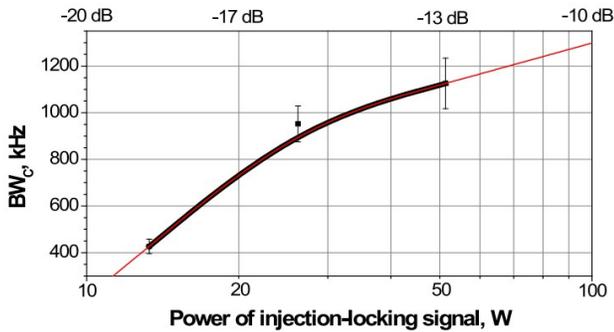

Fig. 10. The admissible bandwidth of control of 2.45 GHz microwave oven magnetrons determined by measured transfer functions characteristics. Black bold line shows the range and results of measurements with B-spline fit; the thin red line shows extrapolation.

The plotted curves indicate that even for 650 MHz magnetron RF sources the bandwidth of control should be ~1 MHz at quite large injected resonant signal.

Estimates of the bandwidth of phase and power control necessary for suppression of microphonics in ADS 1 GeV, SRF, 650 MHz proton driver with the proton beams from 1 to 10 mA are presented in [6].

## 6. On the Life Expectancy of High-Power CW Magnetrons for SRF Accelerators

The high-power CW magnetrons are designed as RF sources for industrial heating, and the lifetime of the tubes is not the first priority as it is required for HEP accelerators. The high-power industrial CW magnetrons use the cathodes made of pure tungsten. The emission properties of the tungsten cathodes are not deteriorated much by electron and ion bombardments, but the latter causes sputtering of the cathode in the magnetron crossed fields. The sputtered cathode material covers interior of a magnetron. This leads to sparks and discharges in a magnetron limiting its lifetime. We considered an analysis of initially developed by Burley Industries Inc. CW, 915 MHz, 10 vanes magnetrons with strapped resonant system vs. failures at power levels of 30, 50 and 75 kW [10]. We performed analysis of impact of ionization of the residual gas in vicinity of the tungsten cathode at 2500 K by a simple model and simulated it by the SRIM program [11]. The yield of cathode neutral atoms [12] caused by accelerated ions of nitrogen hitting the cathode for commercial, the 100 kW, CWM-100L type CW magnetron operating at the magnetic field of 0.238 T at the anode voltage of 22 kV and the emission current density ~0.3 A/cm$^2$ [13] is shown in Fig. 11.

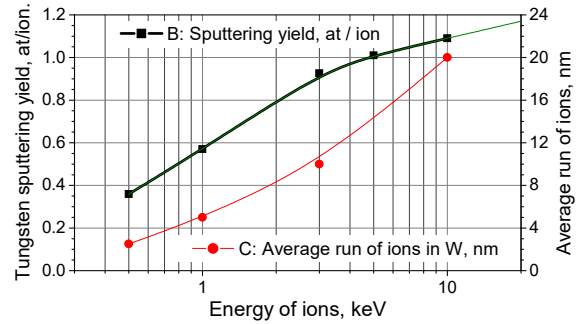

Fig. 11. Dependence of the sputtering yield vs. the ion energy (black dots) and the average penetration depth of ions into the cathode (red dots). Thin olive line shows extrapolation. Simulations imply the single charged nitrogen ions incoming at normal incidence into tungsten at the room temperature..

Based on our model, considering nitrogen residual gas, it follows that for the emitted current of ≈6 A the ion flux is ~2.7·10$^{15}$ ions/s. It corresponds to density of the ion flux on the cathode ~1.4·10$^{14}$ ions/(cm$^2$·s) with the ion energy up to 20 keV. As a result one can expect a loss of tungsten from the cathode ~ 3 g/1000 h. which covers the magnetron interior with average thickness of ~ 40 μm [12]. Note that thickness of the sputtered tungsten may be non-uniform. The sputtering causes arcs and discharges in magnetrons. The atomic content of the residual gas is very important, Fig.12.

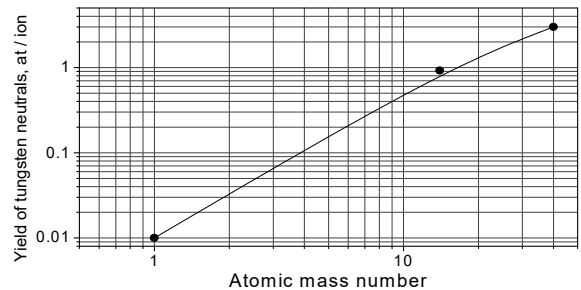

Fig. 12. The yield of the neutrals sputtered from tungsten cathode in dependence on the ions atomic mass numbers A at the ions energy of 3 keV.

The used model shows that a significant improvement of the magnetron vacuum e.g., by the built-in ion pump, may significantly reduce the residual gas pressure. A decrease of the vacuum pressure to about 10$^{-8}$ torr or less would decrease the cathode sputtering resulting in magnetron longevity desired for SRF accelerators.

A lower cathode voltage for the operation in the Stimulated RF generation mode reduces probability of arcs and discharges caused by the cathode sputtering. Also, a reduction of the electron back-stream in the Stimulated RF generation mode somewhat reduces the cathode surface temperature. These phenomena are also useful for increase of the magnetron life time [12]. Using them along improved vacuum properties of magnetrons will help to achieve the desired longevity of the high-power CW magnetrons.

## 6. Summary

The measurements of the spectral density of noise power of CW magnetrons injection-locked by a small resonant signal and operating in the self-excitation mode indicates that the injection-locking bandwidth is narrow as it is predicted by Adler theory. It means low probability of the injection-locked RF generation of the magnetron. In this case, the magnetron generates predominantly noise. Increasing the injection-locking signal to -10 dB reduces the noise power spectral density by ~20 dBc/Hz and indicates a significant increase probability of the injection-locking at $P_{Lock}$ = -10 dB.

Measurements show that the quasi-continuous noise spectra are significantly reduced in operation of the CW magnetrons below the threshold of self-excitation.

The developed Stimulated generation mode eliminates the CW magnetrons noise out at the absence of the injected resonant signal. It means that magnetron in this mode generates the oscillations forced by the injected resonant signal, i.e., the forced coherent oscillations like traditional RF amplifiers.

Efficiency of CW magnetrons in the Stimulated generation mode is higher than efficiency of magnetrons operating in self-excitation modes: free run or driven by a small injection-locking signal, see [2].

The Stimulated generation mode allows adjusting the magnetron power with higher efficiency in a wide range, up to almost the nominal power of the tube.

The Stimulated generation mode is suitable for CW and pulse SRF accelerators. However, the High Voltage pulse modulators shaping pulse anode voltage of magnetrons no need when the Stimulated generation mode is used for pulse SRF accelerators. Stimulated generation mode provides 100% pulse modulation of the magnetron output power at 100% pulse modulation of the resonant injected signal.

The bandwidth of the phase and power control in CW magnetrons operating in the Stimulated generation mode is most suitable for suppression of parasitic modulations in the SRF cavities of accelerators.

A reduced injected resonant forcing signal necessary for the Stimulated generation mode provide two-cascade magnetrons. This allows reduction of power of the injected resonant signal by ≈10 dB. The first CW magnetron with 10% power of the power required from the RF source provides the injected forcing signal for the second, high power CW magnetron. The power control is provided by regulation of the magnetron current in the high power tube. At the mass production the cost of the cascaded RF source will be increased insignificantly.

Analysis of failures of CW, high-power, 915 MHz magnetrons initially developed by Burley Industries Inc. demonstrated that the failures of magnetrons caused by arcs and discharges are significantly increased by increase of the magnetrons anode currents and output power in the self-excitation (free run) mode. A simple model of residual gas ionization in the interaction space of high-power CW magnetrons shows that bombardment of the cathode by accelerated ions leads to magnetron sputtering of neutral cathode atoms (tungsten) covering the inside of the tube.

This limits the life time of the high-power CW magnetrons because of arcs and discharges. A radical method reducing the sputtering is a significant improvement of the magnetron vacuum in the cathode vicinity e.g., by an in-built ion pump. Recently developed technique of Stimulated RF generation of magnetrons is also useful to prolong the high-power CW magnetrons life time since in this case the tube operates at lover anode voltage with reduced electron back-stream.

## Acknowledgment

This manuscript has been authored by collaboration of Muons, Inc-Fermilab under CRADA-FRA-2017-0026 and CRADA NO 2023-0029 between Fermi Research Alliance, LLC Operator of Fermi National Accelerator Laboratory and Muons, Inc. under U.S. Department of Energy Contract No. DE-AC02-07CH11359 with the U.S. Department of Energy, Office of Science, Office of High Energy Physics.

## References

[1] A.C. Dexter, WEIOA04 in Proceed. of LINAC2014, Geneva, Switzerland, (2014).
[2] R.J. Paskuinelli, RF Sources, PIPII XMAS, Feb. 26, 2014.
[3] G. Kazakevich, R. Johnson, V. Lebedev, V. Yakovlev, V. Pavlov, PRAB 21,062001 (2018).
[4] P.L. Kapitza, High Power Electronics, Sov. Phys. Usp. 5, 777 (1963)
[5] L.A. Vainstein and V.A. Solntsev, in Lectures on micro-wave electronics, Moscow Sov. Radio, 1973 (in Russion)
[6] G. Kazakevich, R.P. Johnson, Ya. Derbenev, V. Yakovlev, NIM A 1039 (2022) 167086.
[7] G. Kazakevich, R.P. Johnson, T. Khabiboulline, V. Lebedev, G. Romanov, V. Yakovlev, NIM A 980 (2020) 164366.
[8] G. Kazakevich,V. Lebedev, V. Yakovlev, V. Pavlov, NIM A 839 (2016) 43-51.
[9] G. Kazakevich, "High-Power Magnetron RF Source for Intensity-Frontier Superconducting Linacs", EIC 2014, TUDF1132, (2014), http://appora.fnal.gov/pls/eic14/agenda.full.

[10] G.A. Solomon, "Analisys of the magnetron failure modes versus power level", Proceed. of the International Workshop on Crossed-Field Devices, Ann Arbor, Michigan, 1995.
[11] J.F. Ziegler, J.P. Biersack, M.D. Ziegler, "SRIM-the Stopping and Range of Ions in Matter", Nuclear Instrum. Methods in Phys. Research B 268 (2010) 1818–1823. http://www.srim.org
[12] G. Kazakevich, R.P. Johnson, V. Yakovlev FERMILAB-PUB_24-0085-TD-V (2024)
[13] A.D. Andreev, K.J. Hendricks, Journ. of Microwave Power and Electromagnetic Energy, 44 (2), 2010, pp. 114-124.